\documentclass[aps,physrev,reprint,noeprint]{revtex4-2} 
\usepackage{bm, braket,amsmath,mathtools,comment}
\usepackage{graphicx,color}
\usepackage{hyperref}

\begin{document}
\title{Efficient numerical approach for the simulations of high-power dispersive readout with time-dependent unitary transformation}
\author{Shimpei Goto}
\email[]{goto.las@tmd.ac.jp}
\author{Kazuki Koshino}
\email[]{kazuki.koshino@osamember.org}
\affiliation{College of Liberal Arts and Sciences, Tokyo Medical and Dental University, Ichikawa, Chiba 272-0827, Japan}
\date{\today}
\begin{abstract}
    We develop an efficient numerical approach for simulating the high-power dispersive readout in circuit quantum electrodynamics.
    In the numerical simulations of the high-power readout, a large-amplitude coherent state induced in a cavity is an obstacle because many Fock states are required to describe such a state.
    We remove the large-amplitude coherent state from the numerical simulations by simulating the dynamics in a frame where the amplitude of the coherent state is almost absent.
    Using the developed method, we numerically simulate the high-power dispersive readout of the two-level system and the transmon.
    Our proposed method succeeds in producing reasonable behaviors of the high-power dispersive readout which can be deduced from the photon-number dependence of the cavity frequency: The high-power dispersive readout works in the two-level-system case while it does not work in the transmon case.
\end{abstract}
\maketitle
\section{Introduction\label{sec:introduction}}

Technologies in circuit quantum electrodynamics (cQED) are rapidly developing~\cite{nakamura_coherent_1999,koch_charge-insensitive_2007,schreier_suppressing_2008,yan_tunable_2018,arute_quantum_2019,wu_strong_2021} to realize fault-tolerant quantum computers.
Among these technologies, readout techniques are vital since the final procedure of quantum computations is always the readout of qubit information. 
Moreover, the mid-circuit readout is also essential to execute quantum error corrections~\cite{shor_scheme_1995,kitaev_fault-tolerant_2003,fowler_surface_2012,bluvstein_quantum_2022,livingston_experimental_2022,zhao_realization_2022}.
A fast and reliable readout method is a crucial building block to improving the performance of quantum computers.

Dispersive readout~\cite{blais_cavity_2004,vijay_observation_2011} is a ubiquitous method in cQED because this readout scheme is applicable to any type of qubit.
In the dispersive readout, the qubit-state-dependent frequency shift of a cavity mode can be detected from the reflection or transmission of coherent light input~\cite{krantz_quantum_2019}.
With a naive thought, one can increase the reflected or transmitted signal by using high-power-input light and make the readout time shorter.
However, the working principle of the dispersive readout is based on the perturbative theory~\cite{blais_cavity_2004,bravyi_schriefferwolff_2011} and the cavity photon number can be regarded as the perturbation parameter effectively.
High-power light induces the large cavity photon number, and the perturbation theory would break down.
Furthermore, the situation becomes more complicated when the qubit is implemented with the transmon~\cite{koch_charge-insensitive_2007,schreier_suppressing_2008}.
The transmon can be treated as an anharmonic oscillator whose number of eigenstates is not bounded.
Within the eigenstates in the anharmonic oscillator, only the lowest two states compose computational space.
Previous studies~\cite{khezri_measuring_2016,sank_measurement-induced_2016,lescanne_escape_2019,verney_structural_2019,shillito_dynamics_2022-1,khezri_measurement-induced_2022} have reported that input light can induce transitions to the outer space of the computational basis.
Consequently, back action from high-power input is nontrivial in the dispersive readout.
Toward faster readout, it is necessary to analyze these nontrivial effects.

The high-power coherent light also makes the analysis difficult.
The analysis needs numerical treatment since large cavity photon numbers spoil the perturbative treatment, as already stated.
In numerical approaches, large amplitude coherent states induced by the high-power coherent light disrupt the numerical simulations because many Fock states are required to describe such coherent states.
For instance, \textcite{shillito_dynamics_2022-1} tackle this difficulty by utilizing the processing unit designed for large-scale dense linear-algebra operations~\cite{jouppi_domain-specific_2020}.

In this paper, we propose another approach to deal with the difficulty.
We find a way to obtain the displacement operator which can significantly reduce errors coming from the truncation of bosonic degrees of freedom compared to previously adopted displacement operators~\cite{blais_quantum-information_2007,lescanne_escape_2019,verney_structural_2019}.
The obstruction for the numerical simulations is the large amplitude coherent state in the cavity, and the amplitude of the coherent state can be displaced by the displacement operator.
Since the displacement operator is unitary, the displacement can be regarded as the change of a frame.
Consequently, one can numerically simulate the high-power readout in a frame where the amplitude of the coherent state is always zero.
Following this idea, we develop a method to simulate the dispersive readout in such a frame.
Using the developed method, we simulate the dispersive readout in the two-level-system and the transmon cases.
Compared to the displacement operators adopted in previous studies~\cite{blais_quantum-information_2007,lescanne_escape_2019,verney_structural_2019}, 
the developed displacement can simulate the dispersive readout with less Fock states.
The numerical simulations also show that the dispersive readout works even with high-power input in the two-level-system case.
On the contrary, the simulations suggest that the high-power readout does not work in the transmon case.
This difference can be explained by the photon-number dependencies of the cavity frequency in the two cases, and producing the expected behaviors supports the effectiveness of the proposed method in the simulations of the high-power readout.

The rest of the paper is organized as follows: In Sec.~\ref{sec:method}, we introduce the Hamiltonian and the time-dependent unitary transformation. The derivation of the proposed method is also given in this section. In Sec.~\ref{sec:result}, the results of numerical simulations of high-power dispersive readout in the two-level-system and transmon cases are presented. The summary is given in Sec.~\ref{sec:summary}.

\section{Equation of motion with displacement\label{sec:method}}
\subsection{Transformation of Hamiltonian}
\begin{figure}
    \centering
    \includegraphics[width=0.95\linewidth]{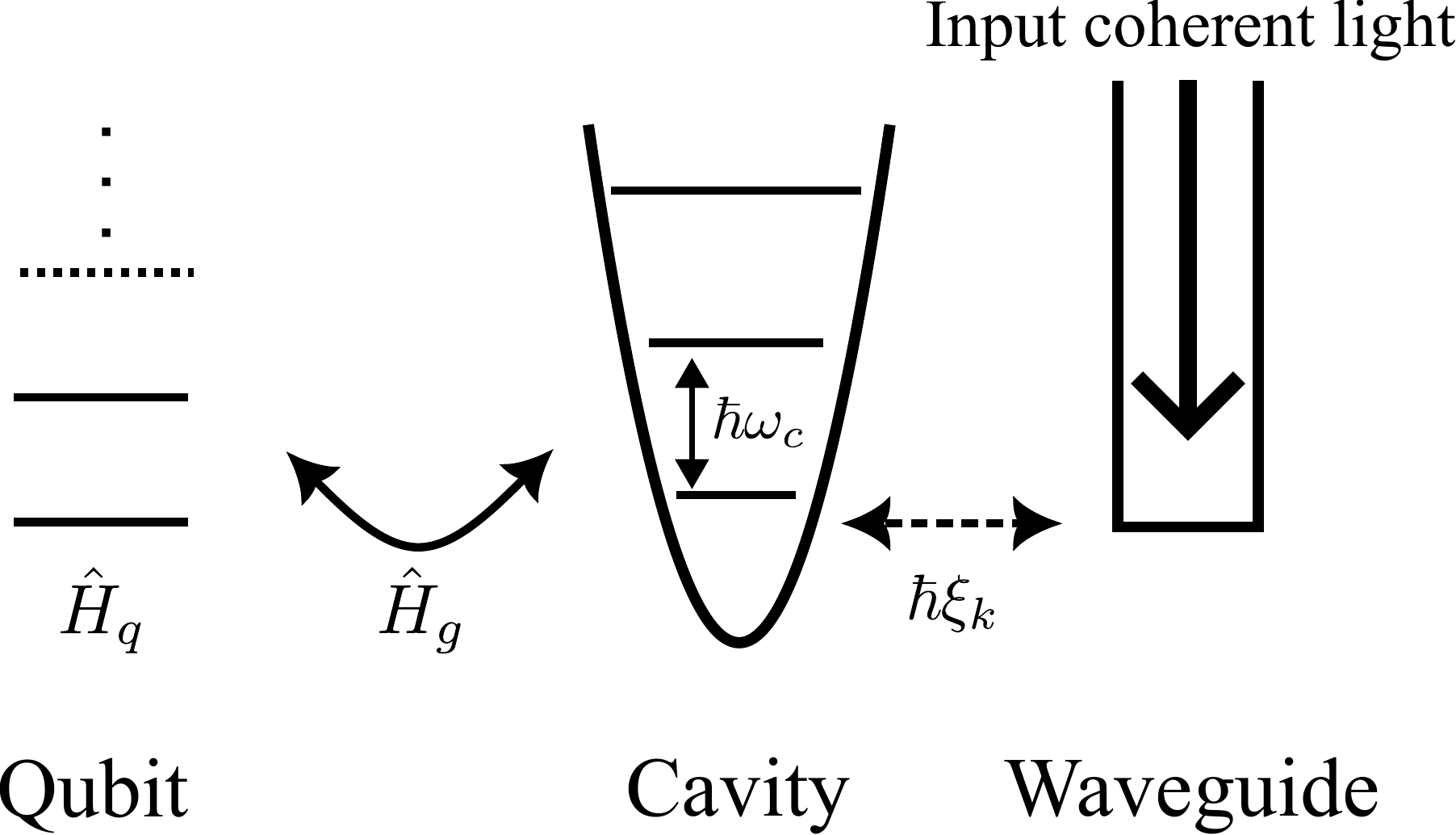}
    \caption{Schematic picture of system we consider in this study. The system is composed of a qubit, a linear cavity, and a semi-infinite waveguide. In this system, the dynamics induced by a coherent input light is considered.\label{fig:system}}
\end{figure}
We consider the dynamics under the Hamiltonian
\begin{align}
    \begin{aligned}
    \hat{H} &= \hat{H}_q + \hat{H}_{g} + \hbar \omega_c \hat{c}^\dagger \hat{c}\\
    &+ \int^\infty_0 dk \left (\hbar v k \hat{b}^\dagger_k \hat{b}_k + \hbar \xi_k\hat{c}^\dagger \hat{b}_k + \hbar \xi^*_k\hat{b}^\dagger_k \hat{c} \right ),
    \end{aligned}
\end{align}
which is depicted in Fig.~\ref{fig:system}.
Here, \(\hat{H}_q\) denotes the Hamiltonian for a component acting as a qubit, \(\hat{H}_{g}\) denotes the interaction between the qubit component and a cavity, \(\omega_c\) is the resonant frequency of the cavity, \(\hat{c}\) (\(\hat{c}^\dagger \)) denotes the bosonic annihilation (creation) operator for the cavity, \(v\) is the velocity of light in a semi-infinite one-dimensional waveguide, \(\hat{b}_k\) (\(\hat{b}^\dagger_k\)) denotes the bosonic annihilation (creation) operator for a mode labeled by a wavenumber \(k\) in the waveguide, and \(\xi_k\) is the coupling between the cavity and the mode \(k\).
For the initial conditions, we assume that the waveguide is not entangled with the other components and that coherent light is injected.
Consequently, an initial state can be represented as
\begin{align}
    \ket{\psi(0)} = \exp\left[\int^\infty_0 dk (f(k)\hat{b}^\dagger_k - f^*(k)\hat{b}_k)\right]\ket{\psi_\mathrm{ini}}_{qc}\ket{0}_w,
\end{align}
where \(f(k)\) is the amplitudes of the input coherent light in the frequency representation, \(\ket{\psi_\mathrm{ini}}_{qc}\) is an initial state of the qubit-cavity system, and \(\ket{0}_w\) denotes the vacuum state of the waveguide.
We also define the Hamiltonian for the qubit and cavity components
\begin{align}
    \hat{H}_{qc} = \hat{H}_q + \hat{H}_{g} + \hbar \omega_c \hat{c}^\dagger \hat{c}.
\end{align}
For labelling the eigenstates of \(\hat{H}_{qc}\), we consider the product states of the eigenstates of \(\hat{H}_q\), \(\ket{p}_q\), and the Fock states \(\ket{i}_c\) of the cavity component which satisfies \(\hat{c}^\dagger \hat{c}\ket{i}_c = i\ket{i}_c\).

At initial, we label a state \(\ket{\tilde{p}, \tilde{0}}_{qc}\) which has the largest overlap with a product state \(\ket{p}_q\ket{0}_c\).
This initial labelling works in the dispersive regime \(|g/(\omega_c - \omega_q)| \ll 1\), where \(\omega_q\) is the resonant frequency of the qubit component.
Starting from \(\ket{\tilde{p}, \tilde{0}}\), states \(\ket{\tilde{p}, \tilde{n}}\) are labelled recursively following the method introduced in Ref.~\cite{shillito_dynamics_2022-1}: A state \(\ket{\tilde{p}, \widetilde{n+1}}\) is characterized as a state which has the largest overlap with the state \(\hat{c}^\dagger \ket{\tilde{p}, \tilde{n}}\).
The eigenenergy of the state \(\ket{\tilde{p}, \tilde{n}}\) is denoted by \(\varepsilon_{p, n}\).
The ground and the first excited states of \(\hat{H}_q\) are denoted by \(\ket{g}_q\) and \(\ket{e}_q\), respectively.

To numerically simulate the dynamics, one has to truncate the infinite Hilbert space of bosonic degrees of freedom to some finite dimension.
In the situation considered in this paper, the input coherent light generates coherent states to bosonic components.
When the input light is strong, the amplitudes of the generated coherent states become large and the dimensions of the truncated local Hilbert spaces should also be large enough to describe these large-amplitude coherent states.
Thus, more computational resources are required for the simulation of the dynamics as input coherent light is stronger.

An approach to reduce the required computational resources is decreasing the amplitudes of the coherent states by the displacement operator
\begin{align}
    \hat{D}(\Delta) = \exp(\Delta\hat{c}^\dagger - \Delta^*\hat{c}).
\end{align}
We introduce the time-dependent unitary operator
\begin{align}
    \hat{U}(t) = \hat{D}^\dagger(\alpha(t))\exp \left [\int^\infty_0 dk (-\beta(k, t)\hat{b}^\dagger_k + \beta^*(k, t)\hat{b}_k)\right ]
\end{align}
and the transformed state
\begin{align}
    \ket{\psi(t)}_U = \hat{U}(t)\ket{\psi(t)}.
\end{align}
By setting \(\beta(k, 0) = f(k)\), the initial state of the waveguide becomes the vacuum state in this frame.
The time evolution of the transformed state \(\ket{\psi(t)}_U\) is governed by the transformed time-dependent Hamiltonian
\begin{align}
    \begin{aligned}
    \hat{H}_U(t) &= \hat{U}(t)\hat{H} \hat{U}^\dagger(t) + i \hbar\left(\frac{d\hat{U}(t)}{dt}\right) \hat{U}^\dagger(t)\\
    &= \hat{H}_q + \hat{D}^\dagger(\alpha(t))\hat{H}_{g}\hat{D}(\alpha(t)) + \hbar \omega_c \hat{c}^\dagger \hat{c}\\
    &+ \int^\infty_0 dk \left (\hbar v k \hat{b}^\dagger_k \hat{b}_k + \hbar \xi_k\hat{c}^\dagger \hat{b}_k + \hbar \xi^*_k\hat{b}^\dagger_k \hat{c} \right )\\
    &+\hbar \left [\hat{c}^\dagger\left (-i\frac{d\alpha(t)}{dt} + \omega_c \alpha(t) + \int^\infty_0 dk \xi_k \beta(k, t)\right ) + \mathrm{H.c.}\right ]\\
    &+\hbar \int^\infty_0 dk \left[ \hat{b}^\dagger_k\left (-i\frac{\partial \beta(k, t)}{\partial t} + v k \beta(k, t) + \xi^*_k \alpha(t) \right ) + \mathrm{H.c.}\right ]\\
    &+C,
    \end{aligned}
\end{align}
where \(C\) denotes c-numbers which are irrelevant to the dynamics and we drop it.
In the transformed frame, the expectation value of an operator \(\hat{O}\) in the original frame can be expressed as
\begin{align}
    \begin{aligned}
    \braket{\psi(t)|\hat{O}|\psi(t)} = \ _{U}\braket{\psi(t)|\hat{U}(t)\hat{O}\hat{U}^\dagger(t)|\psi(t)}_U.
    \end{aligned}
\end{align}
We introduce the notation \(\braket{\hat{O}(t)}_U = \ _{U}\braket{\psi(t)|\hat{O}|\psi(t)}_U\) for later use.

The displacement \(\beta(k, t)\) is chosen so that the condition
\begin{align}
    -i\frac{\partial \beta(k, t)}{\partial t} + vk \beta(k, t) + \xi^*_k \alpha(t) = 0
\end{align}
is fulfilled.
By solving this linear differential equation with the initial condition \(\beta(k, 0) = f(k)\), one can obtain
\begin{align}
    \beta(k, t) = f(k)e^{-ivkt} -i\xi^*_k \int^t_0 d\tau \alpha(\tau)e^{-ivk(t-\tau)}.
\end{align}
Here, we introduce a classical field
\begin{align}
    \mathcal{E}(t) = \int^\infty_0 dk \xi_k f(k) e^{-ivkt}
\end{align}
which corresponds to an external field a cavity feels and the memory function
\begin{align}
    K(t) = \int^\infty_0 dk |\xi_k|^2e^{-ivk t}.
\end{align}
With the introduced quantities, the transformed Hamiltonian can be expressed as
\begin{align}
    \begin{aligned}
        &\hat{H}_U(t) = \hat{H}_q + \hat{D}^\dagger(\alpha(t))\hat{H}_{g}\hat{D}(\alpha(t)) + \hbar \omega_c \hat{c}^\dagger \hat{c}\\
    &+ \int^\infty_0 dk \left (\hbar v k \hat{b}^\dagger_k \hat{b}_k + \hbar \xi_k\hat{c}^\dagger \hat{b}_k + \hbar \xi^*_k\hat{b}^\dagger_k \hat{c} \right )\\
    &+\hbar \Big [\hat{c}^\dagger\left (-i\frac{d\alpha(t)}{dt} + \omega_c \alpha(t) + \mathcal{E}(t) -i \int^t_0 d\tau K(t-\tau)\alpha(\tau)\right )\\
    &\quad \quad + \mathrm{H.c.}\Big ].
    \end{aligned}
    \label{eq:H_trans}
\end{align}

At this point, we introduce two approximations: the extension of the lower limit of the integral for \(k\) in Eq.~\eqref{eq:H_trans} from 0 to \(-\infty \) and ignoring the \(k\)--dependence of \(\xi_k\), i.e., \(\xi_k\) is set to \(\sqrt{\kappa v / (2\pi)}\).
Here, \(\kappa \) represents the decay rate of the cavity.
Under these approximations, the memory function \(K(t)\) is approximated to the delta function, i.e.,
\begin{align}
    K(t) &\simeq \frac{\kappa v}{2\pi} \int^\infty_{-\infty} dk e^{-ivk t}\\
        &= \kappa \delta(t)
\end{align}
and 
\begin{align}
    \int^t_0 d\tau K(t-\tau) \alpha(\tau) &\simeq \kappa \int^t_{0} d\tau \alpha(\tau)\delta(t-\tau)\\
        &= \frac{\kappa}{2} \alpha(t).
\end{align}
In other words, the introduced approximations are equivalent to assuming the Markovian dynamics.

The coefficient of the operator \(\hat{c}^\dagger\) in Eq.~\eqref{eq:H_trans} can be removed by choosing \(\alpha(t)\) as the solution of the linear differential equation
\begin{align}
    \frac{d\alpha(t)}{dt} =  -i \omega_c \alpha(t) -i \mathcal{E}(t) - \frac{\kappa}{2}\alpha(t)
    \label{eq:alpha}
\end{align}
with an initial condition \(\alpha(0) = 0\), and some previous studies have adopted similar choices~\cite{blais_quantum-information_2007,lescanne_escape_2019,verney_structural_2019}.
The solution of this initial value problem is denoted by \(\mathcal{P}(t)\).
For instance, the displacement \(\mathcal{P}(t)\) for the monochromatic field \(\mathcal{E}(t) = Ee^{-i\omega_d t}\) is given as
\begin{align}
\mathcal{P}(t) = \frac{iE\left \{ \frac{\kappa}{2}+i(\omega_d - \omega_c)\right \} }{\frac{\kappa^2}{4}+(\omega_d - \omega_c)^2}(e^{-(i\omega_c + \kappa /2)t}-e^{-i\omega_d t}).
\end{align}
With the choice, the direct driving of the cavity by the field \(\mathcal{E}(t)\) is eliminated from the Hamiltonian, 
\begin{align}
    \begin{aligned}
        &\hat{H}_U(t) = \hat{H}_q + \hat{D}^\dagger(\mathcal{P}(t))\hat{H}_{g}\hat{D}(\mathcal{P}(t)) + \hbar \omega_c \hat{c}^\dagger \hat{c}\\
    &+ \int^\infty_0 dk \left (\hbar v k \hat{b}^\dagger_k \hat{b}_k + \hbar \xi_k\hat{c}^\dagger \hat{b}_k + \hbar \xi^*_k\hat{b}^\dagger_k \hat{c} \right ).
    \end{aligned}
    \label{eq:H_P}
\end{align}
However, the transformed qubit-cavity interaction \(\hat{D}^\dagger(\mathcal{P}(t))\hat{H}_{g}\hat{D}(\mathcal{P}(t))\) can induce a coherent state to the cavity and such coherent states are not taken into considerations in the displacement \(\mathcal{P}(t)\).

\subsection{Equation of motion}
To eliminate a coherent state generated in the cavity from numerical simulations, we derive the Heisenberg equation of motion for an operator acting on the qubit-cavity system \(\hat{s}(t)\)\footnote{In this subsection, \(\hat{D}(\alpha(t))\) denotes the displacement operator in the Heisenberg picture.},
\begin{align}
    \begin{aligned}
    \frac{d}{dt}\hat{s}(t) &= \frac{i}{\hbar}[\hat{H}_q(t) + \hat{D}^\dagger(\alpha(t))\hat{H}_{g}(t)\hat{D}(\alpha(t)), \hat{s}(t)]\\
    &+ i\omega_c [\hat{c}^\dagger(t) \hat{c}(t), \hat{s}(t)]\\
    &+i \sqrt{\frac{\kappa v}{2\pi}}\left([\hat{c}^\dagger(t), \hat{s}(t)]\int^\infty_{-\infty}dk \hat{b}_k(t) + \int^\infty_{-\infty}dk \hat{b}^\dagger_k(t) [\hat{c}(t), \hat{s}(t)]\right)\\
    &+i[\hat{c}^\dagger(t), \hat{s}(t)]\left (-i\frac{d\alpha(t)}{dt} + \omega_c \alpha(t) + \mathcal{E}(t) -i \frac{\kappa}{2}\alpha(t)\right )\\
    &+i[\hat{c}(t), \hat{s}(t)]\left (i\frac{d\alpha^*(t)}{dt} + \omega_c \alpha^*(t) + \mathcal{E}^*(t) +i \frac{\kappa}{2}\alpha^*(t)\right ).
    \end{aligned}
\end{align}
Since the equation of motion for the operator \(\hat{b}_k(t)\) is given as
\begin{align}
    \frac{d}{dt}\hat{b}_k(t) = -i vk \hat{b}_k(t) -i\sqrt{\frac{\kappa v}{2\pi}}\hat{c}(t),
\end{align}
\(\hat{b}_k(t)\) is obtained as
\begin{align}
    \hat{b}_k(t) = \hat{b}_k(0) e^{-ivkt} -i \sqrt{\frac{\kappa v}{2\pi}}\int^t_0d\tau \hat{c}(\tau)e^{ivk(\tau-t)}.
\end{align}
Consequently, one can evaluate the integral
\begin{widetext}
\begin{align}
    \frac{1}{\sqrt{2\pi}}\int^\infty_{-\infty}dk \hat{b}_k(t) &= \frac{1}{\sqrt{2\pi}}\int^\infty_{-\infty}dk\hat{b}_k(0)e^{-ivkt} -i\frac{\sqrt{\kappa v}}{2\pi} \int^\infty_{-\infty}dk\int^t_0 d\tau\hat{c}(\tau) e^{ivk(\tau-t)}\nonumber \\
    &= \frac{1}{\sqrt{2\pi}}\int^\infty_{-\infty}dk\hat{b}_k(0)e^{-ivkt}-i\sqrt{\frac{\kappa}{v}} \int^t_0 d\tau\hat{c}(\tau) \delta(\tau-t)\nonumber \\
    &= \frac{1}{\sqrt{2\pi}}\int^\infty_{-\infty}dk\hat{b}_k(0)e^{-ivkt} -\frac{i}{2}\sqrt{\frac{\kappa}{v}} \hat{c}(t).
\end{align}
\end{widetext}
It should be noted that the first term is the Fourier transform of the operator \(\hat{b}_k(0)\).
Since the waveguide is a vacuum at the initial time in the frame we consider, this term vanishes when one evaluates expectation values in the transformed frame.
Therefore, the equation of motion for the expectation value \(\braket{\hat{s}(t)}_U\) is given as
\begin{widetext}
\begin{align}
    \begin{aligned}
    \frac{d}{dt}\braket{\hat{s}(t)}_U &= \frac{i}{\hbar}\braket{[\hat{H}_q(t) + \hat{D}^\dagger(\alpha(t))\hat{H}_{g}(t)\hat{D}(\alpha(t)), \hat{s}(t)]}_U+ i\omega_c \braket{[\hat{c}^\dagger(t) \hat{c}(t), \hat{s}(t)]}_U + \frac{\kappa}{2}\left(\braket{[\hat{c}^\dagger(t), \hat{s}(t)]\hat{c}(t)}_U - \braket{\hat{c}^\dagger(t) [\hat{c}(t), \hat{s}(t)]}_U\right)\\
    &+i\braket{[\hat{c}^\dagger(t), \hat{s}(t)]}_U\left (-i\frac{d\alpha(t)}{dt} + \omega_c \alpha(t) + \mathcal{E}(t) -i \frac{\kappa}{2}\alpha(t)\right ) +i\braket{[\hat{c}(t), \hat{s}(t)]}_U\left (i\frac{d\alpha^*(t)}{dt} + \omega_c \alpha^*(t) + \mathcal{E}^*(t) +i \frac{\kappa}{2}\alpha^*(t)\right ).
    \end{aligned}
    \label{eq:eom}
\end{align}
\end{widetext}

From this equation of motion, we determine \(\alpha(t)\) to fulfill the condition \(\frac{d}{dt}\braket{\hat{c}(t)}_U = 0\).
By substituting \(\hat{c}(t)\) for \(\hat{s}(t)\), the equation of motion for \(\braket{\hat{c}(t)}_U\) is given as
\begin{align}
    \begin{aligned}
    \frac{d}{dt}\braket{\hat{c}(t)}_U &= \frac{i}{\hbar}\braket{[\hat{D}^\dagger(\alpha(t))\hat{H}_{g}(t)\hat{D}(\alpha(t)), \hat{c}(t)]}_U\\
    &- \left(i\omega_c + \frac{\kappa}{2}\right)(\braket{\hat{c}(t)}_U + \alpha(t)) -i\mathcal{E}(t)\\
    &-\frac{d \alpha(t)}{dt}.
    \end{aligned}
\end{align}
Therefore, the amplitude of the coherent state in the cavity remains its initial value by choosing \(\alpha(t)\) to satisfy the condition
\begin{align}
    \begin{aligned}
    \frac{d\alpha(t)}{dt} &= \frac{i}{\hbar}\braket{[\hat{D}^\dagger(\alpha(t))\hat{H}_{g}(t)\hat{D}(\alpha(t)), \hat{c}(t)]}_U\\
    &- \left(i\omega_c + \frac{\kappa}{2}\right)(\braket{\hat{c}(t)}_U + \alpha(t)) -i\mathcal{E}(t).
    \end{aligned}
    \label{eq:alpha_cond}
\end{align}
We note that an initial value \(\braket{\hat{c}(0)}_U\) can be always set to zero by choosing suitable \(\alpha(0)\).
Consequently, one can simulate the dynamics in the frame where the amplitude of the coherent state \(\braket{\hat{c}(t)}_U\) is always zero.
Since the condition~\eqref{eq:alpha_cond} contains time-dependent expectation values, one has to solve Eqs.~\eqref{eq:eom} and~\eqref{eq:alpha_cond} simultaneously.
With the condition \eqref{eq:alpha_cond}, the equation of motion~\eqref{eq:eom} is rewritten as
\begin{widetext}
\begin{align}
    \begin{aligned}
    \frac{d}{dt}\braket{\hat{s}(t)}_U &= \frac{i}{\hbar}\braket{[\hat{H}_q(t) + \hat{D}^\dagger(\alpha(t))\hat{H}_{g}(t)\hat{D}(\alpha(t)), \hat{s}(t)]}_U+ i\omega_c \braket{[\hat{c}^\dagger(t) \hat{c}(t), \hat{s}(t)]}_U\\
    &+ \frac{\kappa}{2}\left(\braket{[\hat{c}^\dagger(t), \hat{s}(t)]\hat{c}(t)}_U - \braket{\hat{c}^\dagger(t) [\hat{c}(t), \hat{s}(t)]}_U\right)\\
    &+\braket{[\hat{c}^\dagger(t), \hat{s}(t)]}_U\left \{\frac{i}{\hbar}\braket{[\hat{D}^\dagger(\alpha(t))\hat{H}_{g}(t)\hat{D}(\alpha(t)), \hat{c}(t)]}_U - \left(i\omega_c + \frac{\kappa}{2}\right)\braket{\hat{c}(t)}_U\right \}\\
    &-\braket{[\hat{c}(t), \hat{s}(t)]}_U\left \{\frac{i}{\hbar}\braket{[\hat{D}^\dagger(\alpha(t))\hat{H}_{g}(t)\hat{D}(\alpha(t)), \hat{c}^\dagger(t)]}_U + \left(i\omega_c - \frac{\kappa}{2}\right)\braket{\hat{c}^\dagger(t)}_U\right \}.
    \end{aligned}
    \label{eq:eom_nonlinear}
\end{align}
\end{widetext}
In this form, Eqs.~\eqref{eq:alpha_cond} and \eqref{eq:eom_nonlinear} can be regarded as simultaneous ordinary differential equations.
It should be noted that the equation of motion~\eqref{eq:eom_nonlinear} becomes nonlinear by introducing the condition~\eqref{eq:alpha_cond}.
As far as we investigate, an explicit method like the Runge-Kutta method is sufficient to numerically integrate the equations and thus the nonlinearity does not introduce a significant extra cost.
The displacement \(\alpha(t)\) determined from these simultaneous differential equations is denoted by \(\mathcal{Q}(t)\).

In specific, we numerically obtain the dynamics of the expectation values of operators \(\hat{s} = \ket{m}_q\ket{i}_c\bra{n}_q\bra{j}_c\) in the transformed frame.
Here, \(\ket{m}_q\) and \(\ket{n}_q\) are the basis states of the qubit component, and \(\ket{i}_c\) and \(\ket{j}_c\) are the Fock states of the cavity.
For the numerical solver of the simultaneous differential equations, we adopt the Dormand-Prince method~\cite{dormand_family_1980}, which is the fifth-order Runge-Kutta method with an adaptive step size.

\section{Application to dispersive readout\label{sec:result}}
\subsection{Two-level system}
We first demonstrate the performance of our proposed scheme in the two-level-system case, i.e.,
\begin{align}
    \hat{H}_q = \frac{\hbar \omega_q}{2} \hat{Z}
\end{align}
and
\begin{align}
    \hat{H}_{g} = \hbar g \hat{X}(\hat{c}^\dagger + \hat{c}).
\end{align}
Here, \(\hat{X}\) and \(\hat{Z}\) are the Pauli-X and Z operators acting onto the two-level system, respectively, and \(g\) denotes the coupling between the two-level system and the cavity.
In this section, \(\omega_q / \omega_c\) and \(g/\omega_c\) is set to \(0.75\) and \(3.0 \times 10^{-2}\), respectively.
In the parameter region \(|g \sqrt{\braket{\hat{c}^\dagger \hat{c}}}/(\omega_c - \omega_q)| \ll 1\) where the perturbative treatment can be justified, the cavity frequency behaves as \(\omega_c - \chi \hat{Z}\).
Here \(\chi \) is the dispersive shift given by \(g^2/(\omega_c - \omega_q)\)~\cite{blais_cavity_2004,blais_circuit_2021}.
The decay rate of the cavity \(\kappa \) is set to \(2\chi \).
The highest Fock state of the cavity used in numerical simulations is denoted by \(\ket{N_{\max}}\). 

\begin{figure}
    \includegraphics[width=\linewidth]{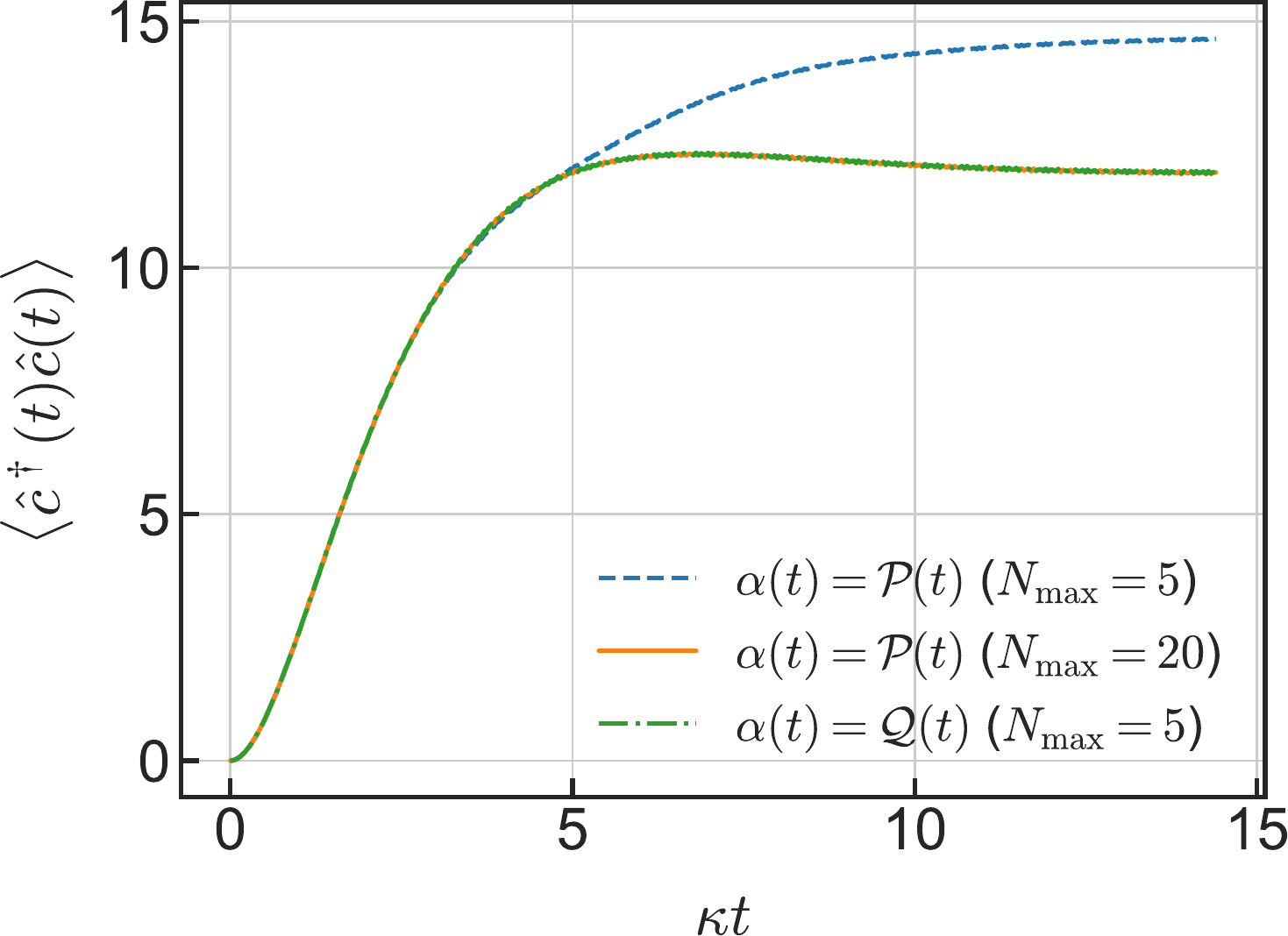}
    \caption{Time evolution of the cavity photon number under a monochromatic drive \(Ee^{-i\omega_c t}\). The displacement \(\mathcal{P}(t)\) is determined by the condition~\eqref{eq:alpha}, and \(\mathcal{Q}(t)\) is determined by our proposed condition~\eqref{eq:alpha_cond}. \(N_{\max}\) is the maximum occupation number of the cavity we set in the simulations. Initially, \(\ket{\psi_\mathrm{ini}}_{qc}\) is set to \(\ket{\tilde{g}, \tilde{0}}\). The parameters used in the simulation are \((\omega_q/\omega_c, g/\omega_c, \kappa/\omega_c, E/\omega_c) = (0.75, \text3.0\times 10^{-2}, 7.2\times 10^{-3}, 1.0\times 10^{-2})\).\label{fig:Nc}}
\end{figure}

To show that our proposed displacement \(\mathcal{Q}(t)\) can describe the dynamics with less \(N_{\max}\) compared to the case with \(\mathcal{P}(t)\),
we evaluate the time evolution of the cavity photon number \(\braket{\hat{c}^\dagger \hat{c}}\) with the monochromatic input field \(\mathcal{E}(t) = Ee^{-i\omega_c t}\).
Figure~\ref{fig:Nc} represents the time evolution of the cavity photon number under the monochromatic drive.
The amplitude of the input field \(E\) is set to \(1.0 \times 10^{-2} \omega_c\) which induces the cavity photon number \(\braket{\hat{c}^\dagger(t)\hat{c}(t)} \sim 10\) in this setting.
An initial state \(\ket{\psi_\mathrm{ini}}_{qc}\) is set to \(\ket{\tilde{g}, \tilde{0}}\).
With the displacement \(\mathcal{P}(t)\), the cavity photon number calculated with \(N_{\max} = 5\) is considerably different from that obtained with \(N_{\max} = 20\) for \(\kappa t \gtrsim 5.0\).
Setting the highest occupation number to \(5\) is insufficient for this dynamics with \(\mathcal{P}(t)\).
In contrast, the calculation with the displacement \(\mathcal{Q}(t)\) and \(N_{\max} = 5\) gives almost the identical cavity photon numbers to those obtained with \(\mathcal{P}(t)\) and \(N_{\max} = 20\).
It should be noted that the dynamics with \(\mathcal{Q}(t)\) can correctly describe the dynamics where the cavity photon number exceeds the highest occupation number \(N_{\max}\).
These results demonstrate the advantage of our proposed displacement \(\mathcal{Q}(t)\) over the displacement \(\mathcal{P}(t)\).

\begin{figure}
    \centering
    \includegraphics[width=\linewidth]{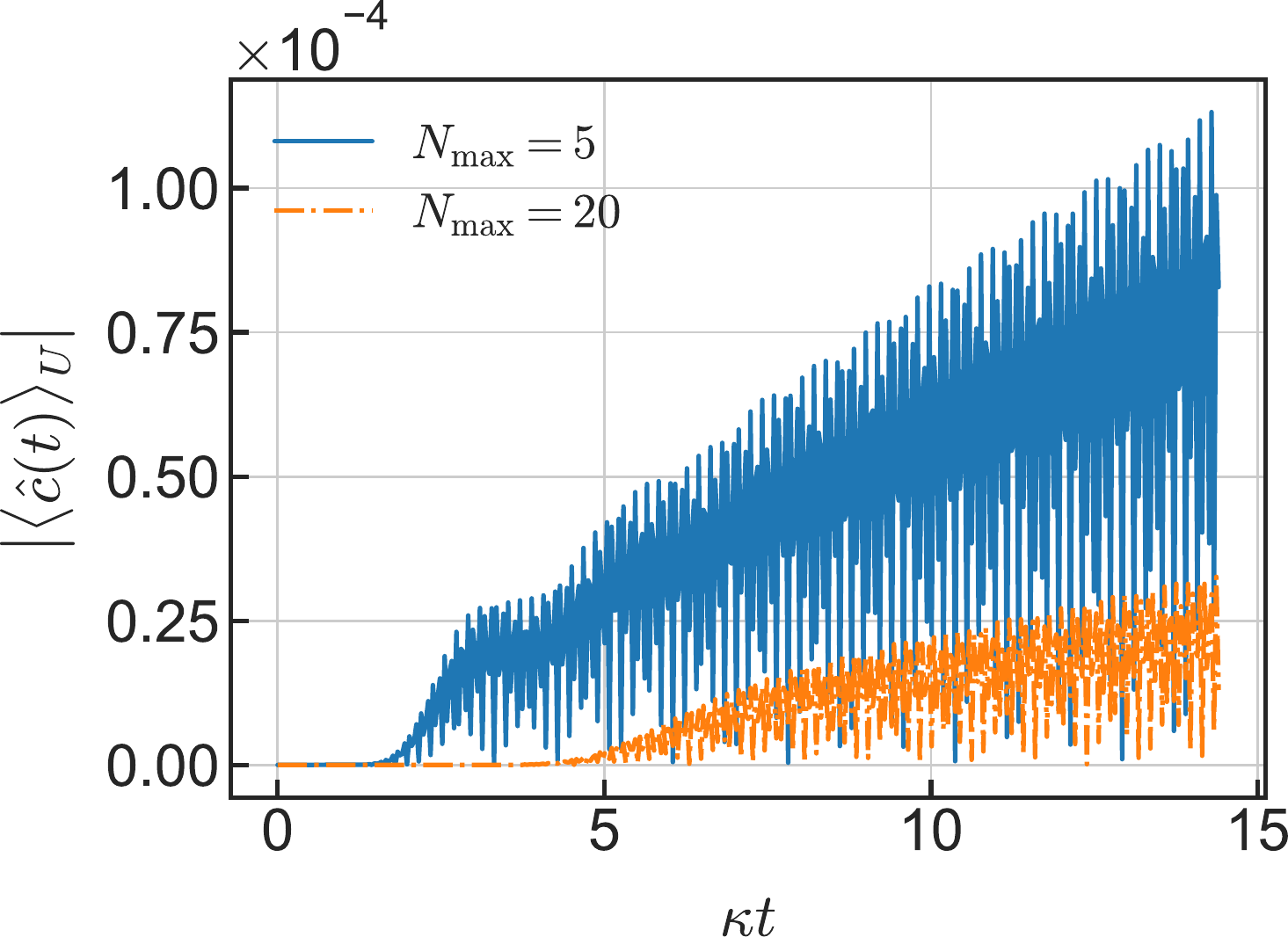}
    \caption{Time evolution of the absolute amplitude of the coherent state generated in the cavity in the transformed frame. The displacement \(\alpha(t)\) is set to \(\mathcal{Q}(t)\). The parameters and the initial state are the same with those in Fig.~\ref{fig:Nc}.\label{fig:Cabs}}
\end{figure}

We design \(\mathcal{Q}(t)\) to eliminate \(\braket{\hat{c}(t)}_U\) from numerical simulations.
Figure~\ref{fig:Cabs} shows the time evolution of the absolute value of \(\braket{\hat{c}(t)}_U\) in the same dynamics presented in Fig.~\ref{fig:Nc}.
With \(N_{\max} = 5\), the absolute values of \(\braket{\hat{c}(t)}_U\) are on the order of \(10^{-5}\).
Although the displacement \(\mathcal{Q}(t)\) works as expected, small but finite values remain. 
Since these values decrease with increasing \(N_{\max}\) up to 20, the small discrepancies from zero would be the results of the truncation of infinite Hilbert space.
From the observation, one can use \(\left | \braket{\hat{c}(t)}_U \right |\) as a measure of the numerical error due to finite \(N_{\max}\).

\begin{figure*}
    \centering
    \includegraphics[width=0.9\linewidth]{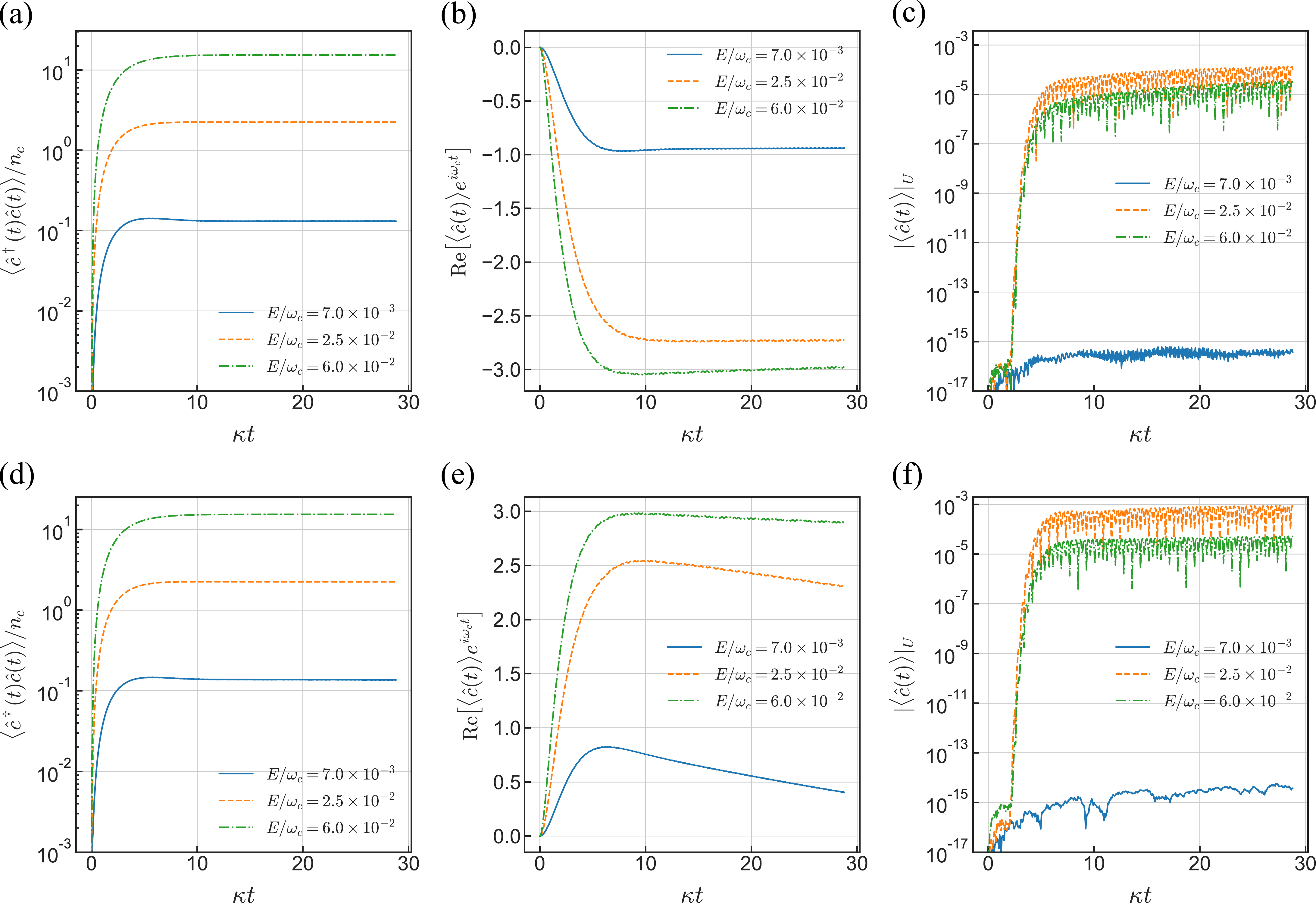}
    \caption{(a),(d) Time evolution of the cavity photon number for different input-field amplitudes. Here, \(n_c\) is the critical photon number given by \({(\omega_q - \omega_c)}^2/4g^2\). The resonant frequency \(\omega_q\) and the coupling \(g\) are the same as those in Fig.~\ref{fig:Nc}. (b), (e) Time evolution of the real part of the cavity amplitude in the rotating frame at the drive frequency starting.  (c), (f) Time evolution of the absolute amplitude of the coherent state in the transformed frame. Initial states are (a-c) \(\ket{\tilde{g}, \tilde{0}}\) and (d-f) \(\ket{\tilde{e}, \tilde{0}}\). The highest occupation number \(N_{\max}\) is set to 30 in the cases with \(E/\omega_c = 7.0 \times 10^{-3}\) and \(2.5 \times 10^{-2}\). For the cases with \(E/\omega_c = 6.0 \times 10^{-2}\), we set \(N_{\max}\) to 50.\label{fig:readout_qubit}}
\end{figure*}

\begin{figure}
    \centering
    \includegraphics[width=\linewidth]{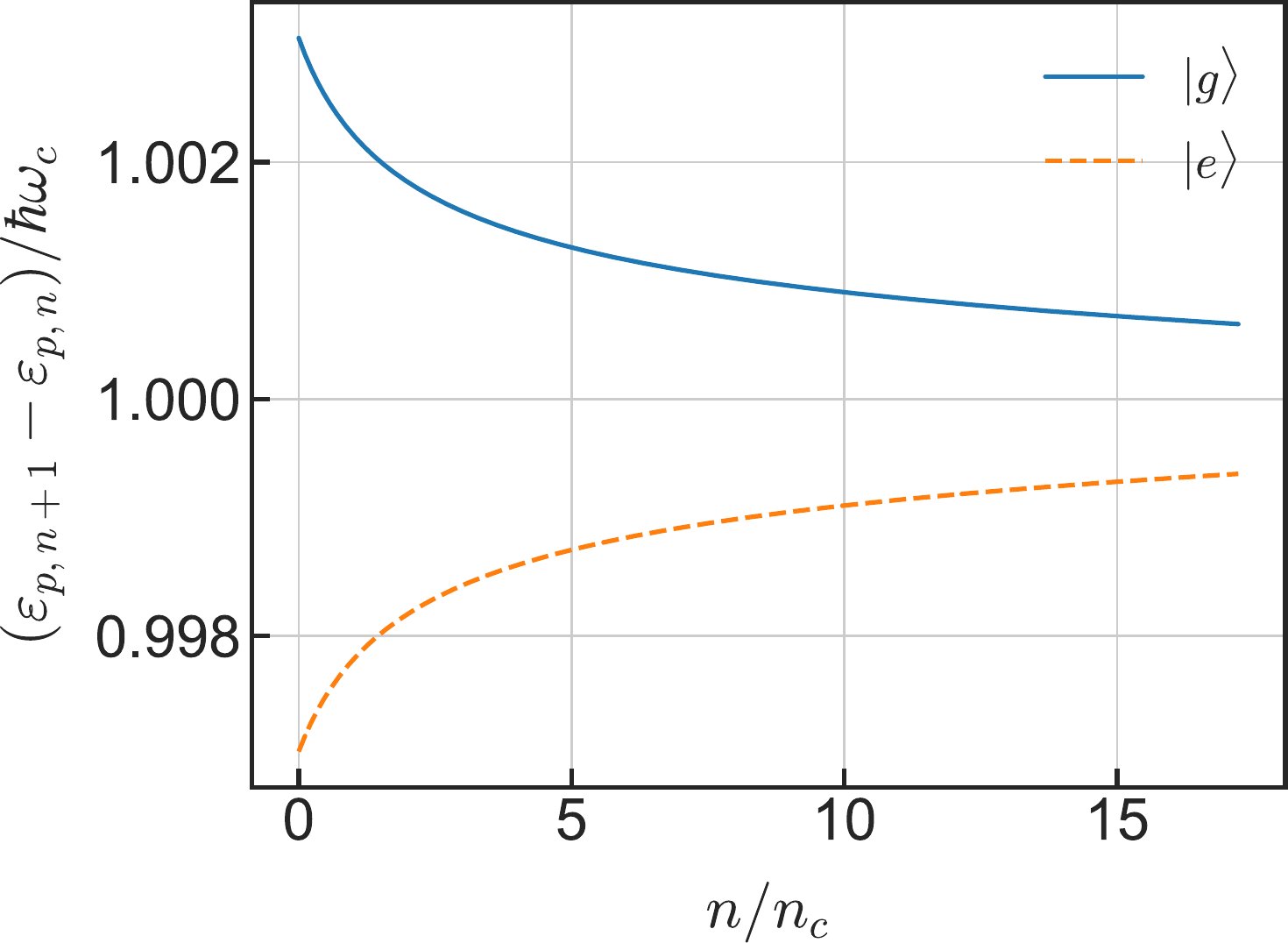}
    \caption{Photon-number dependence of the cavity frequency in the two-level-system case. The resonant frequency \(\omega_q\) and the coupling \(g\) are the same as those in Fig.~\ref{fig:Nc}.\label{fig:qubit_spec}}
\end{figure}

When the drive frequency is tuned to the bare cavity frequency \(\omega_c\), the sign of detuning between the drive frequency and the shifted cavity frequency depends on the qubit state.
The expectation value of one quadrature of a field inside the cavity (the real amplitude \(\braket{\hat{c}^\dagger(t) + \hat{c}(t)}\) in this setting) in the frame rotating at the drive frequency inherits this sign dependence.
In the dispersive readout, the qubit state can be judged from the sign of the quadrature which can be detected with homodyne detection~\cite{blais_cavity_2004,krantz_quantum_2019}.
Figure~\ref{fig:readout_qubit} represents the cavity photon numbers and the real amplitudes of the cavity obtained by the numerical simulations with the displacement \(\mathcal{Q}(t)\).
We consider three cases: The cavity photon number is much smaller than the critical photon number \(n_c = {(\omega_q - \omega_c)}^2/4g^2 \simeq 17.36\) (\(E/\omega_c = 6.0 \times 10^{-3}\)), comparable to \(n_c\) (\(E/\omega_c = 2.5 \times 10^{-2}\)), and much larger than \(n_c\) (\(E/\omega_c = 7.0 \times 10^{-2}\)).
The perturbative treatment is not applicable when the cavity photon number is comparable to or larger than \(n_c\).
Nevertheless, the sign of the real amplitude depends on initial states in all cases.
The dispersive readout works with \(\braket{\hat{c}^\dagger \hat{c}}/n_c \sim \mathcal{O}(10)\) in the two-level system.
This behavior can be understood from the photon-number dependence of the cavity frequency which is given by \(\varepsilon_{p, n+1} - \varepsilon_{p, n}\).
As shown in Fig.~\ref{fig:qubit_spec} , the sign of detuning between the drive frequency \(\omega_c\) and the shifted cavity frequency does not change even in the high occupancy region.
The working principle of the dispersive readout still holds.

\subsection{Transmon}

Next, we apply our proposed scheme to the transmon case~\cite{koch_charge-insensitive_2007,blais_circuit_2021,vool_introduction_2017}, i.e.,
\begin{align}
    \begin{aligned}
    \hat{H}_q &= 4 E_C \sum^\infty_{n=-\infty} {(n-N_g)}^2\ket{n}_q\bra{n}_q\\
    &- \frac{E_J}{2}\sum^\infty_{n=-\infty}(\ket{n}_q\bra{n+1}_q + \ket{n+1}_q\bra{n}_q)
    \end{aligned}
    \label{eq:H_transmon}
\end{align}
and
\begin{align}
    \hat{H}_g = i\hbar g(\hat{c}^\dagger - \hat{c})\sum^\infty_{n=-\infty} (n-N_g)\ket{n}_q\bra{n}_q.
\end{align}
Here, \(E_C\) is the charging energy, \(\ket{n}_q\) denotes the charge basis, \(N_g\) is the offset charge, and \(E_J\) is the Josephson energy.
In numerical simulations, we use \((E_C/\hbar \omega_c, E_J/\hbar \omega_c, g/\omega_c, N_g) = (5.0\times 10^{-2}, 1.6, 3.0 \times 10^{-2}, 0.0)\) so that the energy difference between \(\ket{g}_q\) and \(\ket{e}_q\) is close to that of the two-level-system case.
For the charge basis, we consider states from \(\ket{-10}\) to \(\ket{10}\).
Thus, the dimension of the local Hilbert space for the transmon component is truncated to 21.
The low-lying eigenstates of \(\hat{H}_q\) can be correctly described within the limited Hilbert space.
Specifically, the occupations of \(\ket{-10}\) and \(\ket{10}\) in the eighth excited state are only on the order of \(10^{-12}\). 
Under these settings, the obtained energy difference between \(\ket{g}_q\) and \(\ket{e}_q\) is approximately \(0.7462 \hbar \omega_c\).
The energy difference between \(\ket{e}_q\) and \(\ket{f}_q\) is approximately \(0.6867 \hbar \omega_c\), where \(\ket{f}_q\) is the second excited state of \(\hat{H}_q\).
Consequently, the anharmonicity of this transmon is estimated to be \(-5.95 \times 10^{-2}\hbar \omega_c\).

We evaluate the renormalized cavity frequency and the dispersive shift before the simulation of dynamics.
For the evaluation, we put the energy difference \(\varepsilon_{g, 1} - \varepsilon_{g, 0}\) (\(\varepsilon_{e, 1} - \varepsilon_{e, 0}\)) as \(\hbar\omega^\prime_c + \hbar\chi \) (\(\hbar \omega^\prime_c - \hbar\chi \)).
By numerically diagonalizing \(\hat{H}_{qc}\) with the above parameters, this procedure gives the estimations \(\omega^\prime_c/\omega_c \simeq 1.001975\) and \(\chi/\omega_c \simeq 8.096 \times 10^{-4}\).
These values are of the same order of magnitude with the expressions given by the perturbation theory~\cite{koch_charge-insensitive_2007,blais_circuit_2021}
\begin{align}
    \begin{aligned}
    \omega^\prime_{c,p} - \omega_c &=  \frac{g^2}{\omega_c - \omega_q + E_C/\hbar} \\
    &= 3.0 \times 10^{-3} \omega_c
    \end{aligned}
\end{align}
and
\begin{align}
    \begin{aligned}
    \chi_p &= \frac{g^2E_C/\hbar}{(\omega_c - \omega_q)(\omega_c - \omega_q +E_C/\hbar)} \\
    &= 6.0 \times 10^{-4} \omega_c,
    \end{aligned}
\end{align}
Here, \(\omega^\prime_{c,p}\) and \(\chi_p\) are the renormalized cavity frequency and the dispersive shift given by the perturbation theory, respectively.
Hence, one can adopt the critical photon number based on the perturbation theory~\cite{blais_circuit_2021}
\begin{align}
    \begin{aligned}
    n_c &= \frac{1}{3}\left ( \frac{|\omega_c - \omega_q + E_C/\hbar|^2}{4g^2}- 1\right )\\
    &= 8.0
    \end{aligned}
\end{align}
since only its order of magnitude is relevant.
The decay rate of the cavity \(\kappa \) is set to \(2\chi \) in the following simulations.

\begin{figure}
    \centering
    \includegraphics[width=\linewidth]{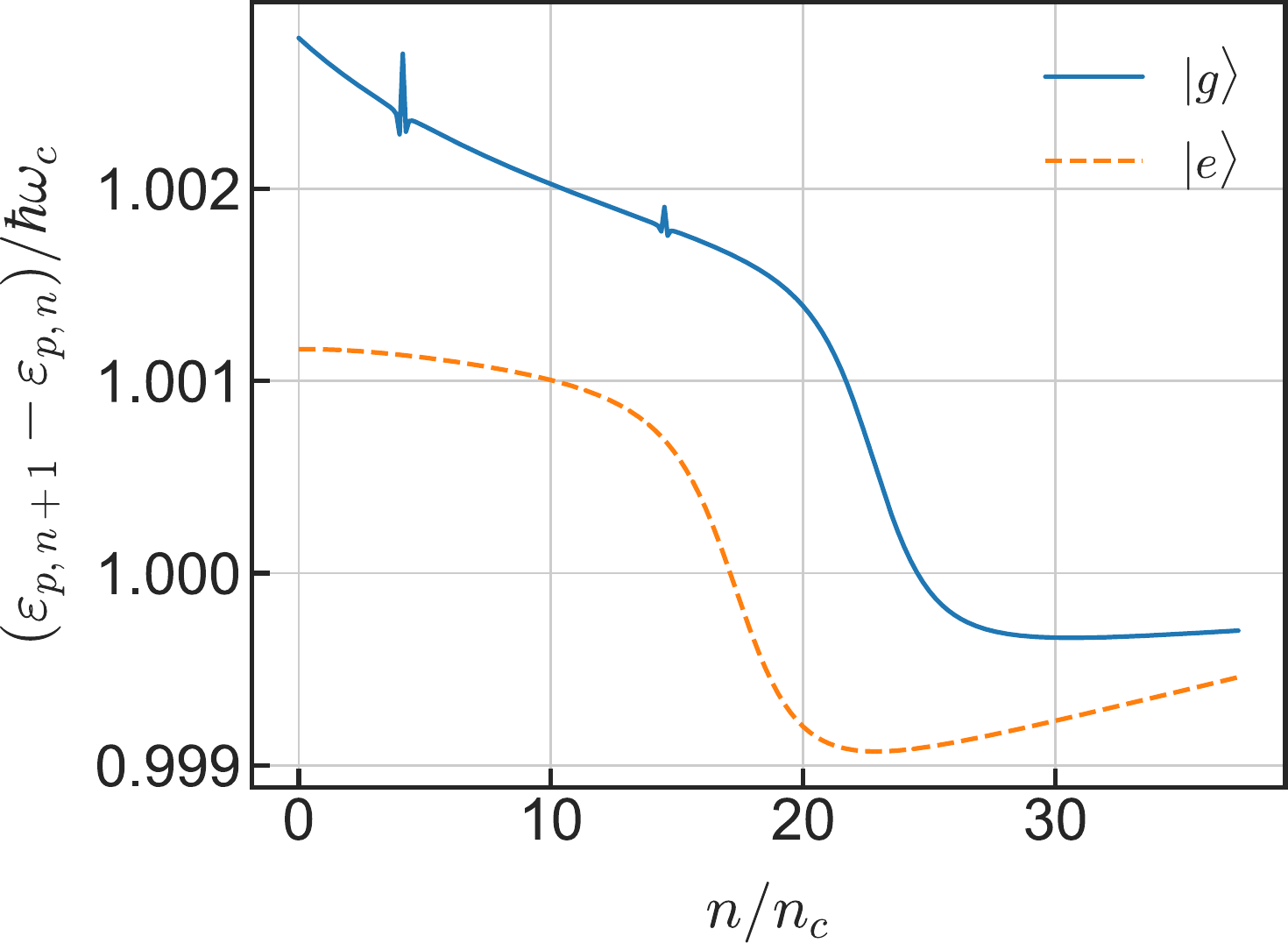}
    \caption{Photon-number dependence of the cavity frequency in the transmon case. For parameters, we use \((E_C/\hbar \omega_c, E_J/\hbar \omega_c, g/\omega_c, N_g) = (5.0\times 10^{-2}, 1.6, 3.0 \times 10^{-2}, 0.0)\).\label{fig:transmon_spec}}
\end{figure}

To determine the drive frequency \(\omega_d\), we investigate the photon-number dependence of the cavity frequency \(\varepsilon_{p, n+1} - \varepsilon_{p, n}\) which is represented in Fig.~\ref{fig:transmon_spec}.
The photon-number dependence in the transmon case is complicated compared with the two-level-system case, and the change of the detuning sign is inevitable with increasing the cavity photon number.
From this photon-number dependence, we set the drive frequency \(\omega_d\) to \(1.0015 \omega_c\).
With the choice, the detuning sign is preserved up to around \(n/n_c \sim 20\).
\begin{figure}
    \centering
    \includegraphics[width=\linewidth]{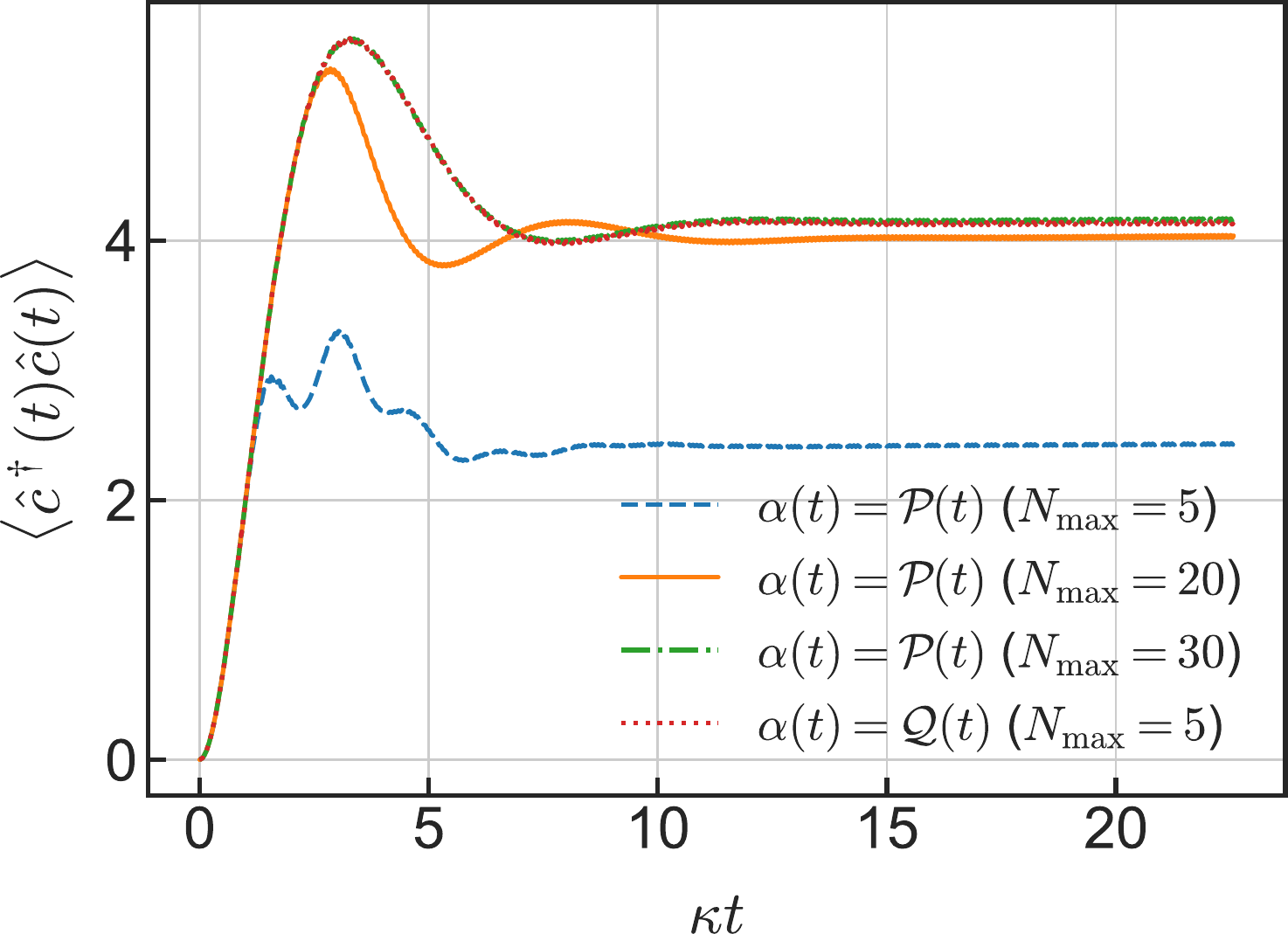}
    \caption{Time evolution of the cavity photon number of the cavity under a monochromatic drive \(Ee^{-i\omega_d t}\) in the transmon case. Initially, \(\ket{\psi_\mathrm{ini}}_{qc}\) is set to \(\ket{\tilde{g}, \tilde{0}}\). The parameters used in the simulation are \((E_C/\hbar \omega_c, E_J/\hbar \omega_c, g/\omega_c, \kappa/\omega_c, N_g, \omega_d/\omega_c) = (5.0 \times 10^{-2}, 1.6, 3.0 \times 10^{-2}, 1.619 \times 10^{-3}, 0.0, 1.0015)\). The amplitude of input field \(E\) is set to \(3.0 \times 10^{-3}\omega_c\).\label{fig:transmon_comp}}
\end{figure}

The advantage of the proposed displacement \(\mathcal{Q}(t)\) can be confirmed in the transmon case as well.
Figure~\ref{fig:transmon_comp} gives the comparison between the displacements \(\mathcal{P}(t)\) and \(\mathcal{Q}(t)\).
Like the two-level-system case, the simulation with the displacement \(\mathcal{Q}(t)\) requires less cavity states compared to the case with the displacement \(\mathcal{P}(t)\) in the transmon case.

\begin{figure*}
    \centering
    \includegraphics[width=0.9\linewidth]{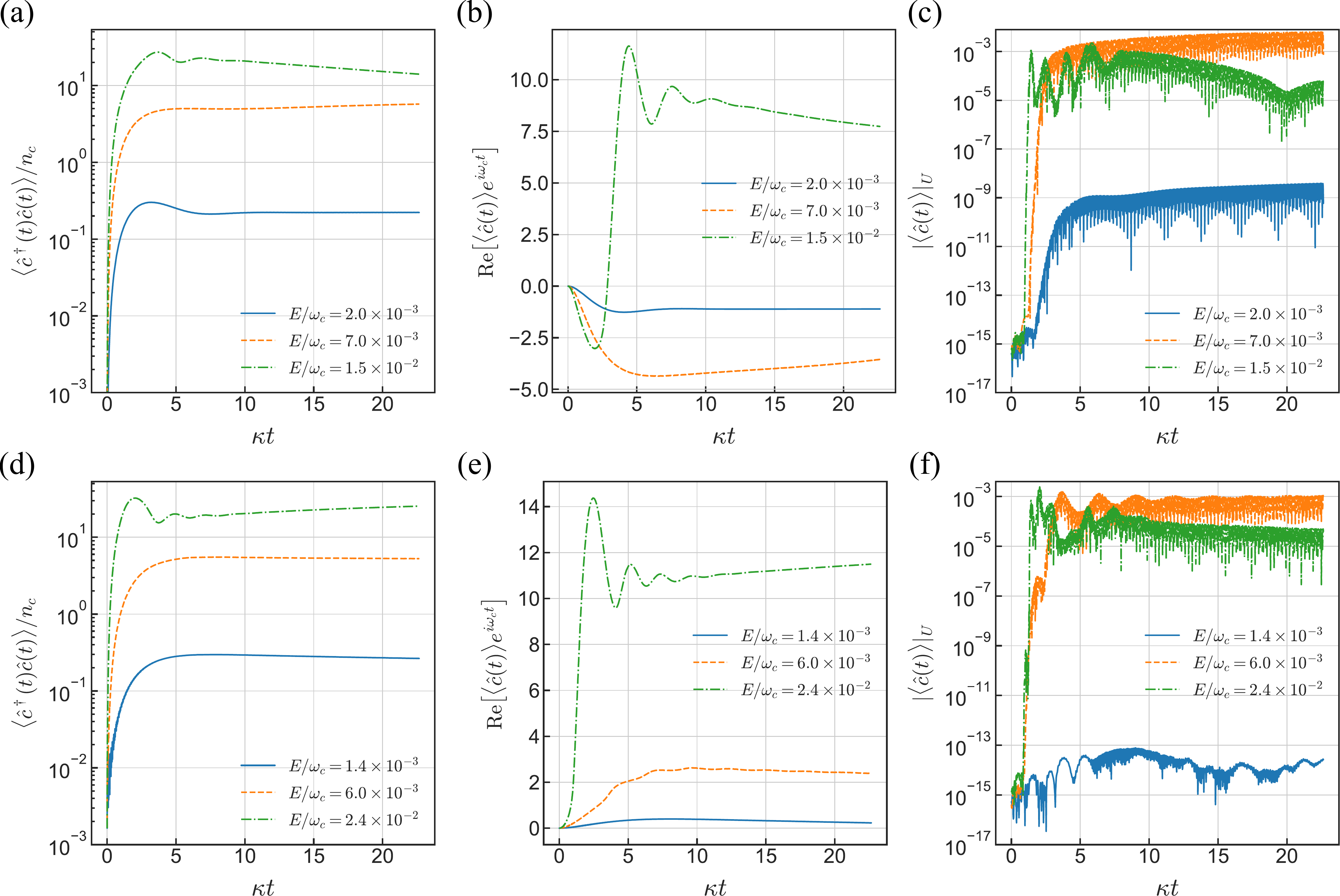}
    \caption{(a),(d) Time evolution of the cavity photon number for different input-field amplitudes in the transmon cases. (b),(e) Time evolution of the real part of the cavity amplitude in the rotating frame at the drive frequency. (c), (f) Time evolution of the absolute amplitude of the coherent state in the transformed frame starting. Initial states are (a-c) \(\ket{\tilde{g}, \tilde{0}}\) and (d-f) \(\ket{\tilde{e}, \tilde{0}}\). The parameters of the system are the same with those in Fig.~\ref{fig:transmon_comp}. The highest occupation number \(N_{\max}\) is set to 20 in the cases with \(E/\omega_c = 1.4 \times 10^{-2}\) and \(2.0 \times 10^{-2}\), 30 in the case with \(E/\omega_c = 6.0\times 10^{-3}\), 60 in the case with \(E/\omega_c = 7.0\times 10^{-3}\), and 100 in the cases with \(E/\omega_c = 1.5 \times 10^{-2}\) and \(2.4 \times 10^{-2}\).\label{fig:readout_transmon}}
\end{figure*}

Figure~\ref{fig:readout_transmon} shows the time evolution of the cavity photon numbers and the real amplitudes of the cavity obtained by the numerical simulations with the displacement \(\mathcal{Q}(t)\) in the transmon case.
Like the two-level-system case, we consider the three cases: The cavity photon number is much less than \(n_c\) (\(E/\omega_c = 1.4 \times 10^{-3}\) and \(2.0 \times 10^{-3}\)), comparable to \(n_c\) (\(E/\omega_c = 6.0 \times 10^{-3}\) and \(7.0 \times 10^{-3}\)), and much larger than \(n_c\) (\(E/\omega_c = 1.5 \times 10^{-2}\) and \(2.4 \times 10^{-2}\)).
For the cases where the cavity photon number is much less than and comparable to \(n_c\), the signs of the real amplitudes reflect whether the initial states are \(\ket{\tilde{g}, \tilde{0}}\) or \(\ket{\tilde{e}, \tilde{0}}\).
However, in the case where the cavity photon number is much larger than \(n_c\), the real amplitude changes its sign around \(\kappa t \sim 3\) when the initial state is \(\ket{\tilde{g}, \tilde{0}}\).
The readout based on the sign of the real amplitude does not work in this case.
The cavity photon number is \(\braket{\hat{c}^\dagger(t) \hat{c}(t)}/n_c \sim 20\) around \(\kappa t \sim 3\) as shown in Fig.~\ref{fig:readout_transmon}(a).
Therefore, the observed break down of the readout scheme in the numerical simulation is consistent with the estimation from the photon-number dependence of the cavity frequency in Fig.~\ref{fig:transmon_spec}.
This consistency supports the availability of the proposed method in the numerical simulations of high-power readout.

\begin{figure}
    \centering
    \includegraphics[width=0.9\linewidth]{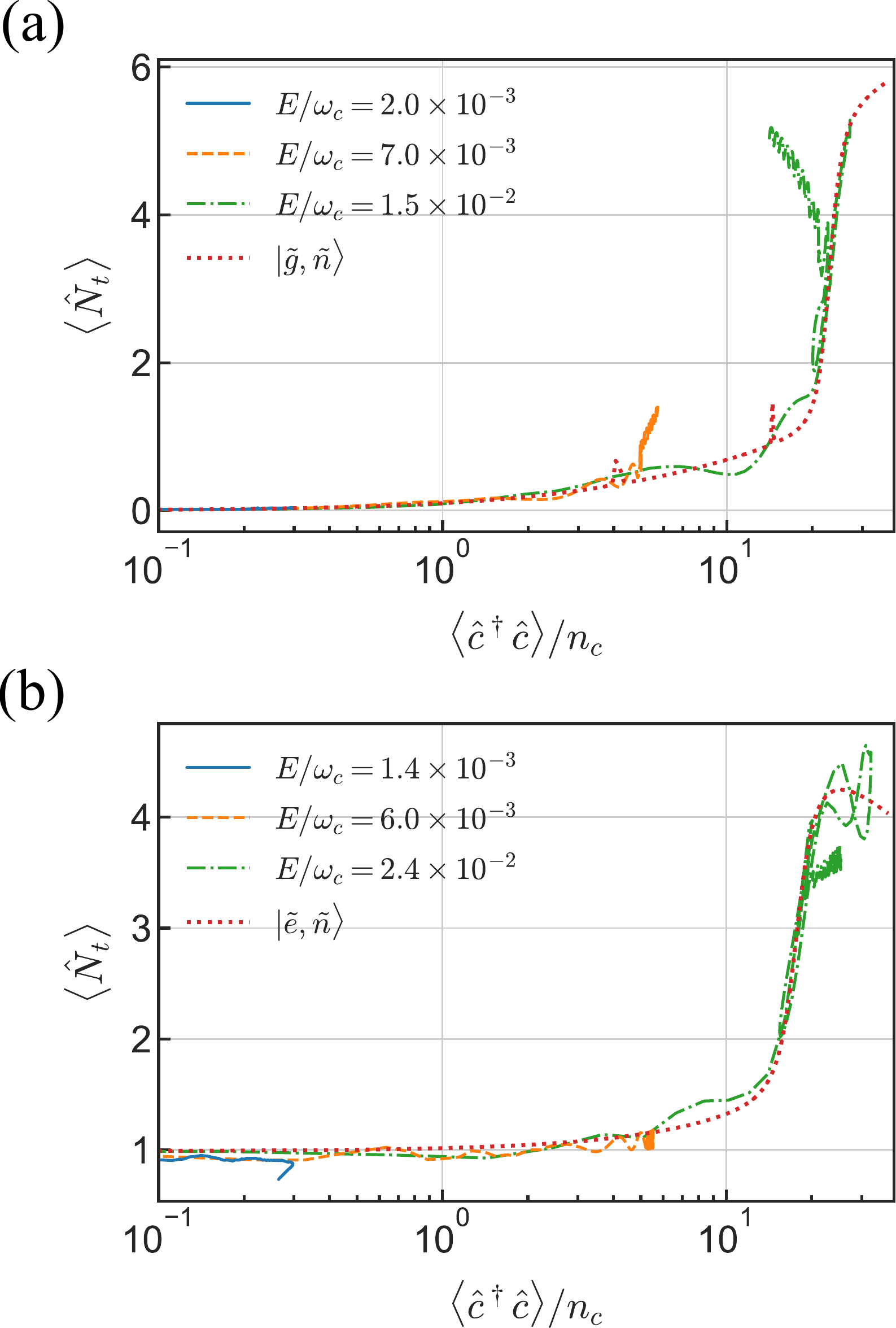}
    \caption{Parametric plot of the transmon occupation number versus the cavity photon number during the dynamics shown in Fig.~\ref{fig:readout_transmon}. Initial states are (a)\(\ket{\tilde{g}, \tilde{0}}\) and (b) \(\ket{\tilde{e}, \tilde{0}}\). Red dotted lines represent the transmon occupation numbers as a function of the cavity photon number obtained from the labeled eigenstates \(\ket{\tilde{g}, \tilde{n}}\) and \(\ket{\tilde{e}, \tilde{n}}\).\label{fig:Nt}}
\end{figure}

\begin{figure}
    \includegraphics[width=0.9\linewidth]{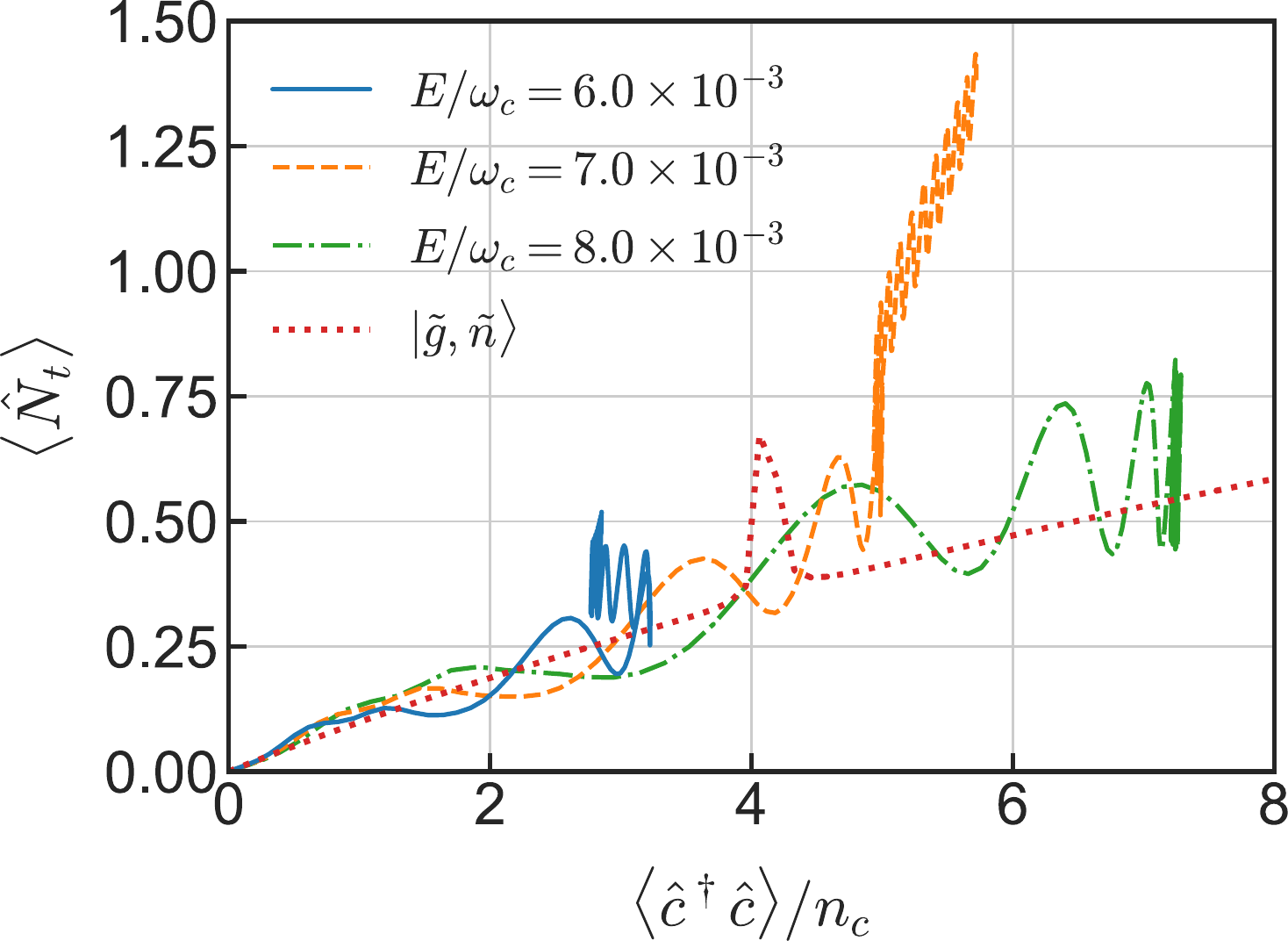}
    \caption{Parametric plot of the transmon occupation number versus the cavity photon number during the dynamics with input-field amplitudes \(E/\omega_c = 6.0\times 10^{-3}\), \(7.0 \times 10^{-3}\), and \(8.0 \times 10^{-3}\). Red dotted line represents the transmon occupation number as a function of the cavity photon number obtained from the labeled eigenstates \(\ket{\tilde{g}, \tilde{n}}\). An initial state is \(\ket{\tilde{g}, \tilde{0}}\). The highest occupation number \(N_\mathrm{max}\) is set to 40 in the cases with \(E/\omega_c = 6.0 \times 10^{-3}\) and \(8.0 \times 10^{-3}\) and 60 in the case with \(E/\omega_c = 7.0 \times 10^{-3}\). The other parameters are the same as those in Fig.~\ref{fig:readout_transmon}.\label{fig:Nt_g}}
\end{figure}

The leakage from the computational space during the readout is an important phenomenon when the qubit component has more than two levels.
To see the leakage in the dynamics simulated in Fig.~\ref{fig:readout_transmon}, we introduce the transmon occupation operator
\begin{align}
    \hat{N}_t = \sum_{l=1} l \ket{l}_q\bra{l}_q,
\end{align}
and evaluate the transmon occupation number \(\braket{\hat{N}_t}\)~\cite{shillito_dynamics_2022-1}.
Here, \(\ket{l}_q\) is the \(l\)--th excited state of the transmon Hamiltonian~\eqref{eq:H_transmon}.
The transmon occupation number gradually increases in the labelled eigenstates \(\ket{\tilde{p}, \tilde{n}}\) with \(n\).
Consequently, noticeably higher transmon occupation number compared to that from the labelled eigenstates can be treated as the sign of the leakage.

Figure~\ref{fig:Nt} represents the parametric plot of the transmon occupation number versus the cavity photon number during the readout dynamics shown in Fig.~\ref{fig:readout_transmon}.
The figure also gives the transmon occupation number as a function of the cavity photon number in the labeled eigenstates for comparison.
For the most part of the dynamics, the parametric plot shows good agreement with the transmon occupation number obtained from the labelled eigenstates.
In contrast, the dynamics starting from \(\ket{\tilde{g}, \tilde{0}}\) state with \(E/\omega_c = 7.0 \times 10^{-3}\) and \(1.5 \times 10^{-2}\) clearly show higher transmon occupation numbers compared to that of the labeled eigenstates.
This behavior can be considered as the sign of the leakage.

The leakage in the case with \(E/\omega_c = 7.0 \times 10^{-3}\) is triggered by the resonance between \(\ket{\tilde{g}, \tilde{n}}\) and the higher excited state around \(n/n_c \sim 4 \). Figure~\ref{fig:Nt_g} gives the parametric plots of the transmon occupation number versus the cavity photon number with some input-field amplitudes close to \(E/\omega_c = 7.0 \times 10^{-3}\).
The bump around \(\braket{\hat{c}^\dagger \hat{c}}/n_c \sim 4\) in the transmon occupation number of the labeled eigenstates comes from the resonance with the higher excited state (the fifth excited state of the transmon Hamiltonian). 
When the cavity photon number stays near the resonant point, the transmon occupation number differs from that of the labeled eigenstates as shown in the case with \(E/\omega_c = 7.0 \times 10^{-3}\).
On the other hand, the case with the larger input-field amplitude \(E/\omega_c = 8.0 \times 10^{-3}\) does not show the noticeable difference because the resonant point is quickly passed in this case.
Consequently, the leakage in the case with \(E/\omega_c = 7.0 \times 10^{-3}\) can be considered as the result from the resonance.
Similar dynamics has been observed in Ref.~\cite{shillito_dynamics_2022-1}.
The proposed method can describe the leakage dynamics during the readout.

\section{Summary\label{sec:summary}}

In this study, we developed an efficient approach to numerically simulate dynamics with a high-power input field.
Our proposed scheme is based on eliminating large-amplitude coherent states from the simulation by the time-dependent displacement operation.
The displacement introduced in this study outperforms that designed to eliminate the direct driving of a cavity in the sense that the dynamics can be reproduced in smaller Hilbert space.

We also applied our proposed scheme for the simulations of the dispersive readout in the two-level-system and transmon cases.
The proposed scheme enables one to access the dispersive readout where the cavity photon number is much larger than the critical photon number with moderate numerical resources.
The obtained numerical results showed that the readout works in the two-level-system case even though the cavity photon number considerably exceeds the critical photon number.
In contrast, the dispersive readout fails in the transmon case when the cavity photon number is much larger than the critical photon number.
This failure can be explained by the photon-number dependence of the cavity frequency, and the numerical results reproduced the the estimation obtained from the cavity frequency.
The numerical simulations also succeeded in describing leakage dynamics.

Although only the two cases, namely the two-level system and transmon, were considered in this study, our proposed displacement can be applied to other devices as long as their Hamiltonian representations are available.
For input fields, we considered only the monochromatic light.
The proposed scheme can treat other input fields, e.g., bichromatic light or short pulses.
Our proposed scheme has a potential impact on evaluating the performance of newly designed quantum devices and optimizing of the shape of input pulses. 

\begin{acknowledgments}
    We thank T. Shitara, K. Sakai, S. Tamate, Y. Tabuchi, M. Tanaka, and T. Yamamoto for fruitful discussions.
    This work was financially supported by JST Moonshot R\&D Grant Numbers JPMJMS2061 and JPMJMS2067.
\end{acknowledgments}
%

\end{document}